\documentclass[aps,prl,twocolumn,showpacs,groupedaddress]{revtex4}
\usepackage[dvips]{graphicx}
\usepackage{amssymb}
\usepackage{amsmath}
\usepackage{color}

\begin{document}

\title{
Experimental determination of the statistics of photons emitted by
a tunnel junction}
\author{Eva Zakka-Bajjani}
\author{J. Dufouleur}
\altaffiliation{Present address: Walter Schottky Institut
Technische Universit\"at M\"unchen Am Coulombwall 3, 85748
Garching, Germany}
\author{N. Coulombel}
\author{P. Roche}
\author{D.C. Glattli}
\altaffiliation{Also at Laboratoire Pierre Aigrain,
D{\'e}partement de Physique de l'Ecole Normale Sup\'erieure, 24
rue Lhomond, 75231 Paris Cedex 05, France}
\author{F. Portier}\email{fabien.portier@cea.fr} \affiliation{Service de Physique de l'Etat Condens{\'e}/IRAMIS/DSM (CNRS URA 2464), CEA Saclay, F-91191 Gif-sur-Yvette,
France}
\date{\today}

\begin{abstract}

We report on a microwave Hanbury-Brown Twiss experiment probing
the statistics of GHz photons emitted by a tunnel junction in the
shot noise regime at low temperature. By measuring the
cross-correlated fluctuations of the occupation numbers of the
photon modes of both detection branches we show that, while the
statistics of electrons is Poissonian, the photons obey chaotic
statistics. This is observed even for low photon occupation number
when the voltage across the junction is close to $h\nu/e$.

\end{abstract}
\pacs{73.23.-b,73.50.Td,42.50.-p,42.50.Ar} \maketitle

What is the statistics of microwave photons radiated by the
current fluctuations of a quantum conductor? When the conductor is
at equilibrium at temperature $T$, the statistics is that of a
black body radiation \cite{Nyquist28PhysRev32p110}. However little
is known about the statistics of the photons emitted in the
non-equilibrium case where the conductor is biased by a voltage
$V\gg k_{B}T/e$ and the current fluctuations are due to quantum
shot noise. An intriguing question is the link between the
statistics of electrons and that of the emitted photons. This
problem has recently attracted theoretical interest and the full
range of photon statistics, from chaotic to non-classical, has
been predicted. The result depends on the competition between the
fermionic and bosonic statistics of electrons and photons
respectively \cite{BeenakkerSchomerus}. On the one hand, the Pauli
principle makes electrons emitted by a contact essentially
noiseless and the current noise only results from electron
scattering with sub-Poissonian statistics. On the other hand
photons emitted by electrons may show bunching effect transforming
their initial statistics from sub-Poissonian to super- Poissonian.
More generally, this rich physics relates to the problem of the
electron Full Counting Statistics \cite{FCS}, as the second moment
of the photon noise directly links to a fourth-moment of the
current fluctuations
\cite{BeenakkerSchomerus,Gabelli04PRL93p056801,LebedevArxiv}.

In this work, we present the first measurements of the statistics
of photons radiated by a quantum conductor in the shot noise
regime. For the simplest quantum conductor studied here, a tunnel
junction, although the statistics of electrons crossing the
conductor is Poissonian, the photon statistics is shown to be
chaotic. This is found even in the regime of vanishing electron
shot noise where the voltage is close to the photon energy ($eV
\gtrsim h \nu$), so that the photon population is small, in
agreement with the prediction of \cite{BeenakkerSchomerus}. As a
by-product, our experimental method based on Hanbury-Brown Twiss
(HBT) microwave photon correlation is found to provide a direct
measurement of the non symmetrized current noise power, called
emission noise. Here, the experiments linearly amplifies the field
amplitude to a classical level and further detects the microwave
power and its fluctuations. This contrasts with experiments
measuring the average power with on-chip quantum detectors such as
quantum dots or superconducting tunnel junctions \cite{onchip}.

The relation between current noise and photon emission can be
understood following Nyquist's approach
\cite{Nyquist28PhysRev32p110}. Consider a conductor of resistance
$R$ connected to a circuit made of a lossless transmission line of
characteristic impedance $Z_{c}$ terminated by a matched resistor.
To simplify, let's assume that $Z_{c} \ll R$. When electrons in
the conductor generate a current fluctuation $I(t)$, a voltage
$V(t)=Z_{c} I(t)$ builds up at the input of the transmission line
exciting an electromagnetic mode which propagates and is finally
absorbed in the resistive load. Introducing the spectral density
of the current fluctuation $S_{I}(\nu)=2
\int_{-\infty}^{+\infty}\langle I(0)I(\tau )\rangle e^{i2\pi \nu
\tau} d \tau $, we can express the electromagnetic power radiated
by the conductor in frequency range $\nu, \nu +d\nu$ as
$dP=Z_{c}S_{I}(\nu) d\nu=N(\nu)h\nu d\nu$, where $N(\nu)$ is the
mean photon population of the electromagnetic mode at frequency
$\nu$. This establishes a direct link between $S_{I}(\nu)$ and
$N(\nu)$. Let us take a further step and consider the (low
frequency) fluctuations $\delta N^2$ of the photon population.
They originate from the intrinsic fluctuations of the current
noise in the conductor (the 'noise' of the noise). However the
problem of the connection between the statistics of the photons
and the electrons is complicated by the bunching effect occurring
when several photons are simultaneously emitted into the
transmission line. The photon distribution emitted by a classical
current was first addressed by Glauber who showed that the photon
statistics is Poissonian \cite{Glauber51PhysRev84p395}. Solving
the same problem in the case of quantum electronic shot noise
requires a model that treats electrons and the detecting
environment on the same quantum footing. Such a treatment was
recently developed by Schomerus and Beenakker
\cite{BeenakkerSchomerus}. In particular they have shown that a
tunnel junction emits photons with chaotic statistics. This occurs
even in the regime of small photon number when the applied voltage
on the junction is close to $h\nu / e$. This is the regime
addressed by our experiments.

\begin{figure}
\centerline{\includegraphics[angle=-90,width=8.5cm,keepaspectratio,clip]{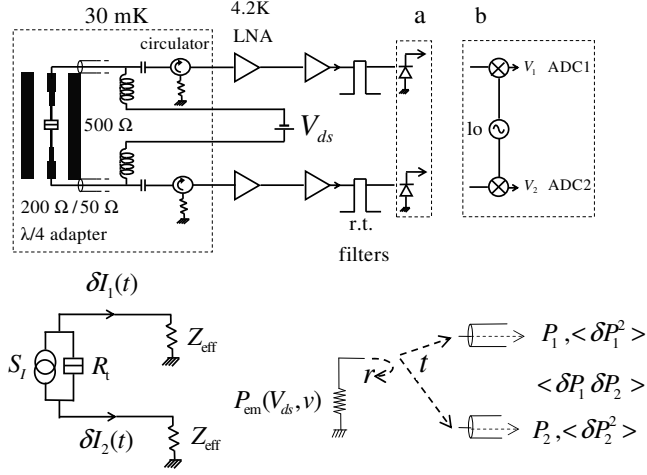}}
\caption{Schematic diagram of the measurement set-up corresponding
to a) detection of the amplified microwave power by quadratic
detectors b) fast digitization of the down-converted current
fluctuations. c) Equivalent microwave circuit.} \label{Schema.fig}
\end{figure}

The experimental set-up, shown in figure \ref{Schema.fig}, is
similar to the one described in \cite{Zakka07PRL99p236803}. An
Al/AlO$_x$/Al tunnel junction of resistance $R_{\mathrm{t}}= 502
\mathrm{\Omega} \pm 1 \mathrm{\Omega}$ was cooled to $\sim 30$ MA
in a dilution fridge. A 0.1 T magnetic field suppresses the Al
superconductivity. The two sides of the tunnel junction are
separately connected to 50 $\Omega$ coaxial transmission lines via
two quarter wave length impedance adapters, raising the effective
input impedance of the detection lines to $Z_{\mathrm{eff}} = 200
\mathrm{\Omega}$ over a one octave bandwidth centered at 6 GHz.
Two rf-circulators, thermalized at mixing chamber temperature
ensure a circuit environment at base temperature.

We note $\delta I_{1,2}$ the fluctuating currents in either
detection branches resulting from the fluctuations of the current
through the tunnel junction. $S_{I_{1}}, S_{I_{2}},
S_{I_{1},I_{2}}$ stand for the autocorrelated and cross-correlated
spectral densities. From the equivalent circuit represented in
figure \ref{Schema.fig}, one easily sees that:

\begin{gather}
S_{I_{1}}(\nu, T, V_{ds})=S_{I_{2}}(\nu, T, V_{ds})=-
S_{I_{1}I_{2}}(\nu, T, V_{ds})\nonumber \\ =
\displaystyle{\left(\frac{R_{\mathrm{t}}}{2 Z_{\mathrm{eff}}+
R_{\mathrm{t}}}\right)^2} S_{I}(\nu, T, V_{ds})
,\label{deltaI1deltaI2}
\end{gather}

\noindent where the two first equalities result from current
conservation. The noise power detected in each detection line in a
frequency range $\Delta \nu$ reads

\begin{equation}
P_{1,2} = Z_{\mathrm{eff}}  \overline{\delta I_{1,2}^2} =
 \frac{ 4 Z_{\mathrm{eff}}R_{\mathrm{t}}}{(2
Z_{\mathrm{eff}}+ R_{\mathrm{t}})^2} P_{\mathrm{em}}
,\label{emittedpower.eq}
\end{equation}
\noindent where $P_{\mathrm{em}}=R_{\mathrm{t}}  S_{I}(\nu, T,
V_{ds}) \Delta \nu/4$ is the emitted power. This can be expressed
as an excess noise temperature $\Delta T_{n
1,2}=P_{1,2}/[k_{\mathrm{B}} \Delta \nu_{1,2}]$.

The two emitted signals are then amplified by two cryogenic Low
Noise Amplifiers.
 Up to a calculable gain factor, the detected noise power
contains the weak excess noise $\Delta T_{n 1,2}$ on top of a
large additional noise generated by the cryogenic amplifiers $T_{n
1,2} \simeq 5 \mathrm{K}$. After further room temperature
amplification and eventually narrow bandpass filtering, current
fluctuations are detected using three alternative techniques.
First (fig. 1.a), we implemented the measurement scheme described
in \cite{Zakka07PRL99p236803}, using two calibrated quadratic
detectors whose output voltage is proportional to noise power.
Secondly (fig. 1.b), current fluctuations are digitized, after
down conversion achieved by mixing with a suited local microwave
signal, using an AP240 Acqiris Acquisition Card able to sample at
1Gsample/s. A quantitative comparison with the well established
first method has qualified this new method. The third method,
dedicated to the study of photon noise, is a hybridization of the
two previous ones: the outputs of the two quadratic detectors are
digitized to perform the photon HBT cross and auto correlations
fluctuations of $P_{1,2}$.

\emph{First experiment: mean photon occupation number.} We measure
the increase in noise temperature due do the photon emission by
shot noise, as a function of $V_{ds}$ and the measuring frequency
$\nu$, using the quadratic detectors. In order to remove the
background noise of the amplifiers, we measure the excess noise,
$\Delta S_{I_{1,2}}(\nu, T, V_{ds})=S_{I_{1,2}}(\nu, T,
V_{ds})-S_{I_{1,2}}(\nu, T, 0)$. Practically, this is done by
applying a 93Hz $0$-$V_{ds}$ square-wave bias voltage on the
sample through the DC input of a bias-Tee, and detecting the first
harmonic of the square-wave noise response of the detectors using
lock-in techniques. The results are quite similar to the one
reported in ref.\cite{Zakka07PRL99p236803}, and lead to an
electron temperature $T_e \sim 70$ MA. Although $T_e$ is
significantly higher than the mixing chamber temperature, it is
low enough to make the thermal population of photons negligible in
the 4-8 GHz frequency range, where all our measurements are done.
In the high bias limit  ($e V_{ds} \gg k_{\mathrm{B}} T, h \nu$),
$\Delta S_I \sim 2 e I$. Equation \ref{emittedpower.eq} then
yields an excess noise temperature $\Delta T_{n 1,2}=e  V_{ds}/4
k_\mathrm{B}$ in both detection branches in the case of ideal
coupling $R_{\mathrm{t}} = 2 Z_{\mathrm{eff}}$. In practice, we
get excess noise temperatures $\sim 2$dB lower than expected from
the independently measured attenuation of the various microwave
components connecting the sample to both amplifiers. A $\sim 100
\mathrm{fF}$ capacitance for the junction, shunting part of the
microwave signal, accounts for this discrepancy. This value is
quite reasonable given the area of our tunnel junction (1.4
$\mu$m$^2$).

\emph{Second experiment: auto and cross-correlated electronic
noise.} We record the current fluctuations with the acquisition
card using a 5 ns sampling time, chosen large enough to avoid any
correlation between successive points, thus maximizing the
effective bandwidth. Here again, we eliminate background noise and
parasitic correlation between the two inputs of the acquisition
card by measuring excess fluctuations
. $\Delta \overline{\delta
V_1^2}$ and $\Delta \overline{\delta V_2^2}$ are proportional to
the excess noise power:

\begin{equation*}
\Delta \overline{\delta V_{1,2}^2} = G_{1,2} Z_0 P_{1,2} = G_{1,2}
Z_0 Z_{\mathrm{eff}} \Delta S_{I_{1,2}} \Delta \nu,
\end{equation*}

\noindent where $Z_0 = 50 \ \mathrm{\Omega}$ is the input
impedance of the acquisition card, $\Delta \nu$ is the bandpass of
the two filters, centered around the same frequency $\nu$, and
$G_{1,2}$ stands for the gain of chain 1,2. The benefit of this
method is that it also gives access to the cross-correlation term:

\begin{equation}
\Delta \overline{\delta V_1 \delta V_2} = \sqrt{G_1 G_2} Z_0
Z_{\mathrm{eff}}\int_{\Delta \nu} \cos(2 \pi \nu \tau) \Delta
S_{I_{1}I_{2}}d \nu
\end{equation}

\noindent where $\tau$ is the difference of propagation time of
electromagnetic waves between the sample and detectors 1 and 2.
One easily gets:

\begin{equation}
\begin{array}{lcl}
\Delta \overline{\delta V_1  \delta
V_2}_{\mathrm{norm}}&=&\frac{\Delta \overline{\delta V_1  \delta
V_2}}{\sqrt{\Delta \overline{\delta V_1^2} \Delta \overline{\delta
V_2^2}}}=- \mathrm{sinc}(\pi\Delta \nu \tau)\cos(2 \pi \nu \tau)
\\ &\simeq& - \cos(2 \pi \nu \tau) \ \ \mathrm{for} \ \Delta\nu
\tau \ll 1.
\end{array}
\label{V1V2.eq}
\end{equation}

\begin{figure}
\centerline{\includegraphics[angle=-90,width=8.5cm,keepaspectratio,clip]{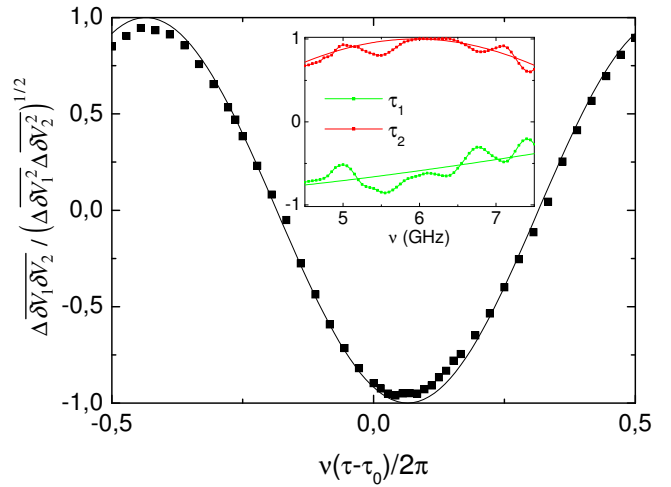}}
\caption{Normalized cross-correlation spectrum $\Delta
\overline{\delta V_1  \delta V_2}_{\mathrm{norm}}$ measured at
$\nu$=6 GHz $\pm$ 50MHz as a function of $\nu \Delta \tau$. The
solid line represents the sinusoidal prediction of eq.
\ref{V1V2.eq}. Inset: $\Delta \overline{\delta V_1  \delta
V_2}_{\mathrm{norm}}$ averaged over 200 MHz, as a function of
$\nu$ for two values of $\tau$. The solid lines correspond to
equation \ref{V1V2.eq} for $\tau=$83 ps (upper line) and $\tau=$12
ps (lower line).} \label{deltaI1deltaI2.fig}
\end{figure}

\noindent Equation \ref{V1V2.eq} expresses the anticorrelation of
current fluctuations $\delta I_1$ and $\delta I_2$,
modified by the phase difference of the microwave signals induced
by $\tau$. This is illustrated in Figure \ref{deltaI1deltaI2.fig} for $\nu=6$ GHz with $\Delta \nu$=100
MHz. Here, $\tau$ is varied using two calibrated phase shifters
inserted in both detection lines, around a value $\tau_0$ which is
\textit{a priori} not known.

As photons emitted at times differing by more than $\sim 1/\Delta
\nu$ do not show correlations, one needs to minimize $\tau$ before
measuring the cross-correlated power fluctuations. As $\Delta \nu
\ll \nu$, the most sensitive way to do so is to ensure that
$\Delta \overline{\delta V_1 \delta V_2}_{\mathrm{norm}}$ is
constant for various values of $\nu$. The result of such an
adjustment is shown in the inset of figure 2. Although small
parasitic microwaves reflections introduce extra modulations,
$\Delta \overline{\delta V_1  \delta V_2}_{\mathrm{norm}}$ doesn't
change sign for $ 4 \mathrm{GHz} \le \nu \le 8 \mathrm{GHz}$. This
implies that $\tau \le 125$ps, so that $\Delta \nu \tau \ll 1$ and
the delay between the two lines doesn't affect the power
correlations.

\emph{Third experiment: auto and cross-correlated photon noise.}
We now adopt the original HBT set-up \cite{Hanbury56Nature177p27},
using quadratic detectors connected to the digitizer. This gives
access to $S_{P^{\mathrm{out}}_1P^{\mathrm{out}}_1}$ and
$S_{P^{\mathrm{out}}_2P^{\mathrm{out}}_2}$, and to the
cross-correlation spectrum
$S_{P^{\mathrm{out}}_1P^{\mathrm{out}}_2}$. In the case of a
tunnel junction consisting of many weakly transmitted electronic
modes, one would expect \cite{Glauber63PhysRev131p2766} the
emitted photons to follow a negative binomial distribution, as the
emitted power results from the incoherent superposition of a large
number of sources (the electronic modes of the tunnel junction).
This is confirmed by the rigorous treatment of ref.
\cite{BeenakkerSchomerus}, with corrections corresponding to
emission of two photons at the same energy by the same electron.
In the case of a low impedance (compared to the resistance quantum
$h/e^2$), these corrections are small and one expects the photons
emitted by the shot noise power of a tunnel junction to have the
same counting statistics as thermal photons, although their origin
is quite different and the frequency dependence of photon
occupation number $N(\nu)$ doesn't correspond to a Bose-Einstein
thermal distribution.
 The cross-correlated power
fluctuations are then expected to be positive, reflecting the
bosonic nature of the emitted photons, and proportional to the
products of the power emitted in both detection branches. The
autocorrelated power fluctuations are expected to be enhanced by
the contribution of the noise of the amplifiers
\cite{Caves82PRD26p1817}. Let's note $\Delta\nu_{\mathrm{min}}$
($\Delta\nu_{\mathrm{max}}$) the smaller (the bigger) of the
bandpasses, both centered around the same frequency $\nu$. One
thus expects:

\begin{equation}
\begin{array}{lcl} S_{P^{\mathrm{out}}_1P^{\mathrm{out}}_1}&=&2   \Delta
\nu_1 \left[ G_1^2 k_{\mathrm{B}}^2  (T_{n1} + \Delta T_{n1})^2 +
\right.
\\ &&\left.\ \ \ \ \ \ \ \ G_1 h \nu k_{\mathrm{B}} (T_{n1}+ \Delta T_{n1})\right]
\\ &\simeq& 2   \Delta
\nu_1 G_1^2 k_{\mathrm{B}}^2  (T_{n1} + \Delta T_{n1})^2\\
S_{P^{\mathrm{out}}_1P^{\mathrm{out}}_2}&=&2 G_1 G_2
\Delta\nu_{\mathrm{min}} k_{\mathrm{B}}^2 \Delta T_{n1} \Delta
T_{n2},
\end{array}
\label{SPP.eq}
\end{equation}

\begin{figure}
\centerline{\includegraphics[angle=-90,width=8.5cm,keepaspectratio,clip]{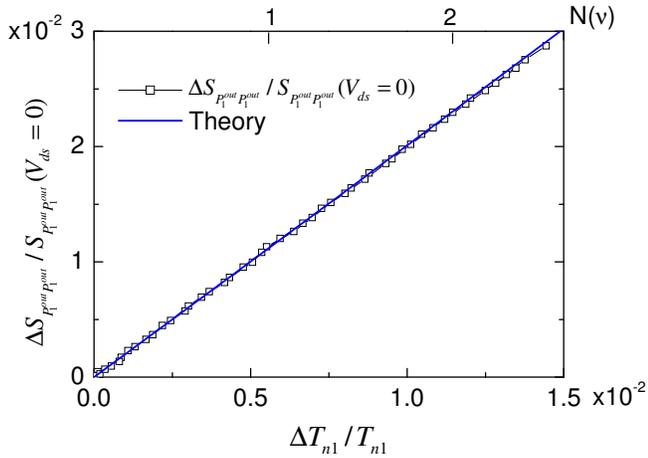}}
\caption{Normalized auto-correlation spectrum of power
fluctuations vs normalized excess noise temperature. Symbols:
experimental data. Line: Theoretical prediction. The top axis
gives the corresponding $N(\nu)$.} \label{deltaP1deltaP1.fig}
\end{figure}

\begin{figure}
\centerline{\includegraphics[angle=-90,width=8.5cm,keepaspectratio,clip]{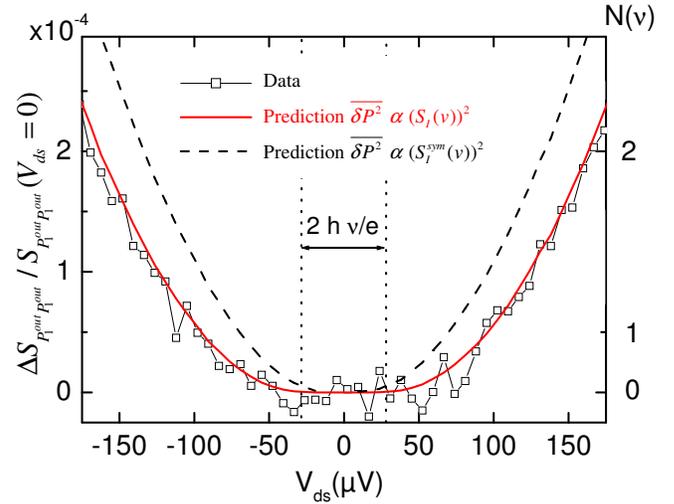}}
\caption{Open squares: Normalized cross-correlation spectrum of
power fluctuations. Solid line: theoretical prediction, assuming
$S_{PP} \propto S_{I}^2$. Dashed line theoretical prediction,
assuming $S_{PP} \propto S_{I,\mathrm{sym}}^2$. The right axis
gives the corresponding $N(\nu)$. The dotted line represent the
onset of finite frequency excess shot noise
\cite{onchip,Zakka07PRL99p236803,eVhnu}.}
\label{deltaP1deltaP2.fig}
\end{figure}

\noindent Equation \ref{SPP.eq} shows the benefit of HBT
cross-correlation to study photon statistics, as it suppresses the
term due to the mixing of the input power with the input noise of
the amplifier. In order to get rid of imperfectly
known gains and attenuation we normalize the excess
autocorrelated power fluctuations by their zero voltage bias values and the cross-correlated power fluctuations by their
geometric mean. Figure \ref{deltaP1deltaP1.fig} represents the excess
power fluctuations spectrum $\Delta
S_{P^{\mathrm{out}}_1P^{\mathrm{out}}_1}$, normalized by the zero
bias value $S_{P^{\mathrm{out}}_1P^{\mathrm{out}}_1} (V_{ds}=0)$
as a function of $\Delta T_{n 1}/ T_{n 1}$, measured at $\nu=6.6\
\mathrm{GHz} \pm 115 \  \mathrm{MHZ}$. The solid line represents
the theoretical prediction:
\begin{equation*}
\frac{\Delta
S_{P^{\mathrm{out}}_1P^{\mathrm{out}}_1}}{\left[S_{P^{\mathrm{out}}_1P^{\mathrm{out}}_1}\right]_{
V_{ds}=0}}=\left(\displaystyle{\frac{\Delta T_{n 1}}{T_{n
1}}}\right)^2 + 2 \displaystyle{\frac{\Delta T_{n 1}}{T_{n 1}}},
\end{equation*}

\noindent which agrees with the experimental observations within
0.5\%. Measurements over the entire 4-8 GHz frequency range show
similar agreement, with a maximum systematic deviation of roughly
3\%. As shown by figure \ref{deltaP1deltaP2.fig}, the
cross-correlated power fluctuations are positive, showing the
bosonic character of the emitted excitations. On a quantitative
level, one should observe:
\begin{equation*}
\frac{\Delta
S_{P^{\mathrm{out}}_1P^{\mathrm{out}}_2}}{\left[S_{P^{\mathrm{out}}_1P^{\mathrm{out}}_1}
S_{P^{\mathrm{out}}_2P^{\mathrm{out}}_2}
\right]_{V_{ds}=0}^{1/2}}=\frac{\Delta T_{n 1}}{T_{n
1}}\frac{\Delta T_{n 2}}{T_{n 2}}\sqrt{\frac{\Delta
\nu_{\mathrm{min}}}{\Delta \nu_{\mathrm{max}}}},
\end{equation*}
\noindent   Figure \ref{deltaP1deltaP2.fig} shows both the
cross-correlated power fluctuations, normalized to the equilibrium
auto-correlated power fluctuations, and the product of the excess
noise temperatures normalized by
$(\Delta\nu_{\mathrm{min}}/\Delta\nu_{\mathrm{max}})^{1/2}$. They
are found to coincide within 4\%. As the 'photon noise' is related
to a fourth order-correlator of the electronic current
\cite{BeenakkerSchomerus,Gabelli04PRL93p056801}, this constitutes,
to the best of our knowledge, the first measurement of such a
correlator in the quantum regime. It shows that the emitted
radiation remains chaotic, even when the occupation number of the
emitted photon modes tends to zero ($eV_{ds}, k_{\mathrm{B}} T \ll
h \nu$).

We would like to add a short note to the question of the
symmetrization of the current correlator involved in this shot
noise measurement. The treatment of Beenakker and
Schomerus\cite{BeenakkerSchomerus} yields power fluctuations
proportional to the square of the electronic emission noise
density \cite{nonsym} $S_{I}(\nu)=2
\int_{-\infty}^{+\infty}\langle I(0)I(\tau )\rangle e^{i2\pi \nu
\tau} d \tau $. Using a symmetrized correlator
$S_{I,\mathrm{sym}}=\frac{1}{2}(S_{I}(\nu)+ S_{I}(-\nu))$
increases the emitted power per unit bandwidth by $h \nu/2$. As
the difference doesn't depend on bias voltage, excess noise
measurements cannot distinguish between the two definitions.
However, the quadratic dependence of the power fluctuations with
the emitted power allows to distinguish them.  As shown by figure
\ref{deltaP1deltaP2.fig}, assuming power fluctuations $S_{PP}
\propto S_{I,\mathrm{sym}}^2$ yields a prediction $\Delta S_{PP}
\propto [(\Delta S_{I})^2+ 4 G_{\mathrm{sample}} h \nu \Delta
S_{I}]$, which is incompatible with our observations.

In conclusion, we have performed the first experiment probing the
statistical properties of photons emitted by a phase coherent
 conductor. The data are found in perfect agreement with
the predictions of Beenakker and Schomerus, showing that a biased
low impedance tunnel junction emits a chaotic radiation. The
cross-correlated power fluctuations are found to be proportional
to the square of the emission electronic noise density. This opens
the way to the investigation of the statistical properties of
photons emitted by mesoscopic conductors where electronic
correlations might have a stronger impact, such as quantum point
contacts, for which it is expected that the sub-Poissonian
statistics of electronic shot-noise could be 'imprinted' on the
corresponding emitted photons \cite{BeenakkerSchomerus,
LebedevArxiv}.

\begin{acknowledgments}
 It's a pleasure to acknowledge precious help from Q.
Le Masne, P. Bertet and D. Vion with the sample fabrication. We
greatly benefited from help during the measurements from B.
Dubost. This work was supported by the ANR contract 2e-BQT, and by
the C'Nano Idf contract QPC-SinPS.
\end{acknowledgments}

\end{document}